%% file: 2011_CH_emission_paper.tex
\documentclass[12pt]{iopart}
\usepackage{graphicx}
\usepackage{multirow}
\usepackage{verbatim} 
\usepackage{wasysym}
\DeclareTextSymbol{\reg}{T1}{174}

\begin{document}
\title[Dissociative recombination and CH emission] {Dissociative recombination and electron-impact de-excitation in CH photon emission under ITER divertor-relevant plasma conditions}
\author{G.A. van Swaaij$^1$, K. Bystrov$^1$, D. Borodin$^2$, A. Kirschner$^2$, L.B. van der Vegt$^1$, G.J. van Rooij$^1$, G. De Temmerman$^1$, W.J. Goedheer$^1$}
\address{$^1$ FOM Institute DIFFER, previously FOM Institute Rijnhuizen, Association EURATOM-FOM, Partner in the Trilateral Euregio Cluster, PO. Box 1207, 3430 BE Nieuwegein, the Netherlands}
\address{$^2$ Institute for Energy and Climate Research - Plasma Physics, Forschungszentrum J\"ulich GmbH, Association EURATOM-FZJ, Partner in the Trilateral Euregio Cluster, 52425, J\"ulich, Germany}
\ead{G.A.vanSwaaij@differ.nl}
\begin{abstract}
\input{abstract}
\end{abstract}
\submitto{\PPCF}
\maketitle

\section{Introduction}
\input{introduction.tex}

\section{CH$_4$ puffing experiments}
\input{puffing.tex}

\section{Simulation details}
\input{simulation.tex}

\section{Results}
\input{results.tex}

\section{CH A-level excitation probabilities of the different dissociative recombination channels}\label{sec:excitation}
\input{exc_prob.tex}

\section{Discussion}\label{sec:discussion}
\input{discussion.tex}

\section{Conclusions}
\input{conclusions}

\section*{Statement of Provenance}
This is an author-created, un-copyedited version of an article accepted for publication in Plasma Physics and Controlled Fusion. IOP Publishing Ltd is not responsible for any errors or omissions in this version of the manuscript or any version derived from it. The definitive publisher-authenticated version will be available online at http://iopscience.iop.org/0741-3335/.
% TODO: when DOI is assigned, make it "The definitive publisher-authenticated version is available online at [INSERT DOI]".

\ack
This work, supported by the European Communities under the contract of Association between EURATOM/FOM,  was carried out within the framework of the European Fusion Programme with financial support from NWO. The views and opinions expressed herein do not necessarily reflect those of the European Commission. The authors would like to thank D. W\"underlich for the useful discussion on the de-excitation of CH, and D. Reiter and R. K. Janev for useful discussions on the dissociation rates.

\section*{References}

\input{refs.tex}
\end{document}

%% file: abstract.tex
For understanding carbon erosion and redeposition in nuclear fusion devices, it is important to understand the transport and chemical break-up of hydrocarbon molecules in edge plasmas, often diagnosed by emission of the CH A$^2\Delta$-X$^2\Pi$ Ger\"o band around 430 nm. The CH A-level can be excited either by electron-impact or by dissociative recombination (D.R.) of hydrocarbon ions. These processes were included in the 3D Monte Carlo impurity transport code ERO. A series of methane injection experiments was performed in the high-density, low-temperature linear plasma generator Pilot-PSI, and simulated emission intensity profiles were benchmarked against these experiments. It was confirmed that excitation by D.R. dominates at $T_e$ $<$ 1.5 eV. The results indicate that the fraction of D.R. events that lead to a CH radical in the A-level and consequent photon emission is at least 10\%. Additionally, quenching of the excited CH radicals by electron impact de-excitation was included in the modeling. This quenching is shown to be significant: depending on the electron density, it reduces the effective CH emission by a factor of 1.4 at n$_e=1.3*10^{20}$ m$^{-3}$, to 2.8 at n$_e=9.3*10^{20}$ m$^{-3}$. Its inclusion significantly improved agreement between experiment and modeling.

%% file: introduction.tex
Carbon has a long history as the plasma-facing material (PFM) of choice for the walls, limiters, and divertors of tokamaks, including divertor plates of ITER. However, upon bombardment with hydrogenic atoms and ions, methane and other hydrocarbon molecules are formed, which can easily leave the PFM surface and enter the edge plasma. This chemical erosion is very efficient even at low ion impact energies {\cite{roth2005}}. At the strike-points of the ITER divertor, where ion fluxes are maximal, electron temperatures ($T_e$) are expected to be around 1-10 eV, much lower than in current tokamaks. The expected electron density ($n_e$) is approximately $10^{20}$ m$^{-3}$, which is much higher than in present tokamaks {\cite{Federici2003JNM_B2EIRENE}}, leading to substantial chemical erosion. The hydrocarbons ejected that way from the PFM will be transported, ionised and dissociated in the edge plasma. They will often be redeposited at other locations, where they may retain unacceptably large amounts of radioactive tritium {\cite{Federici2001Review,temmerman2009}}. Both erosion and tritium retention may limit the availability of next-generation carbon-clad tokamaks and are therefore of utmost importance {\cite{roth2009PWIissues}}.

A common diagnostic for chemical erosion of carbon is the optical emission of the CH radical, in particular the emission around 430 nm due to the A$^2\Delta \rightarrow $X$^2\Pi$ transition (the Ger\"o band). Computer simulations are often benchmarked to these measurements, using the assumption that the A$^2\Delta$ level is primarily excited by inelastic electron-impact (E.I.) collisions. However, the rate of electron impact excitation decreases for lower electron temperatures. Therefore, when electron temperatures are around ITER-divertor-relevant ($T_e < $\ 3 eV) levels, this is no longer necessarily the dominant excitation mechanism of the A$^2\Delta$ level. The exothermic dissociative recombination (D.R.) reactions shown in \tref{DR_reactions_list} can be expected to produce, amongst others, excited CH radicals in the A-level; this process may dominate at low electron temperatures. As previous methane injection experiments \cite{westerhout2010} in the Pilot-PSI linear plasma generator \cite{vanrooij2007_Pilot} have shown, the absolute Ger\"o band emission efficiency stays almost constant when the electron temperature is decreased from 1.2 eV down to 0.1 eV. The cross sections for D.R. are only weakly dependent on temperature; therefore those results suggested that D.R. is the dominant excitation mechanism. This was also seen in expanding thermal plasma \cite{beulens1994Emission}, in the divertor simulator MAP-II \cite{kobayashi2003RotPop}, in a magnetically enhanced capacitive RF discharge \cite{avtaeva2007}, and in plasma spraying \cite{dorier2009DRInSpraying}. Understanding of the photon yield from D.R. is essential for interpretation of CH plumes (i.e. emission patterns), which would give confidence in predictive modeling for ITER plasma-surface interaction and tritium retention \cite{kirschner2007, kirschner2009JNM, kirschner2009PhysScr}.

The importance of D.R. in those plasma conditions was investigated by comparing results of the Monte Carlo impurity transport simulation code ERO {\cite{kirschner2000}} with experiments in the linear plasma generator Pilot-PSI \cite{westerhout2007PhysScr}. The first simulations of hydrocarbon transport in Pilot-PSI have been described elsewhere \cite{borodin2010}; the present paper is a continuation of those modeling efforts. A number of modifications were made to ERO in order to improve its suitability to Pilot-PSI conditions. Several methane puffing experiments were analyzed, including two newly presented experiments. It was investigated whether the combination of excitation by electron impact and dissociative recombination into excited levels can satisfactorily explain the observed photon emission intensities in each of these experiments. De-excitation of the CH A-level by electron impact was included in the modeling, and shown to be an important process at high densities.

\subsection*{Hydrocarbon dissociation and excitation mechanisms at $T_e < 2$ eV}
At low electron temperatures, rates for ionisation and dissociative excitation by electron-impact rapidly go down due to the threshold energies involved. The hydrocarbon (methane) molecules entering the plasma are then predominantly dissociated by a chain of charge exchange reactions, followed by dissociative recombination {\cite{reiter2009}}. Due to the low fraction of hydrocarbon impurities relative to the hydrogen concentration, the reactions between two hydrocarbon molecules can be neglected. Charge exchange (CX) is the dominant mechanism for ionization of the neutral molecules:
\begin{equation}
\mathrm{CH}_{(x)} + \mathrm{H}^+ \rightarrow \mathrm{CH}_{(x-y)}^+ + ( \mathrm{H}, \mathrm{H}_2) \mathrm{\hspace{20pt} (y \leq 1)}.
\end{equation}
There are a number of different dissociative recombination channels that each hydrocarbon ion can undergo. For the methane family (CH$_x^+$), four of these reactions can produce CH radicals, relevant for the Ger\"o band emission. They are listed in \tref{DR_reactions_list}. Each of these reactions is exothermic, and the CH$_2^+$ D.R., CH$_3^+$ D.R. (a), and CH$_4^+$ D.R. reactions are each able to overcome the 2.88 eV electronic excitation threshold of the CH radical. Thus, they should be considered as potential populating processes of the CH(A) level. The CH$_3^+$ D.R. (b) reaction can not excite the A-level.

The probability that each particular reaction actually excites the A$^2\Delta$ level is not known beforehand. There are various other energetically accessible excited states (ie. B$^2\Sigma^-$, C$^2\Sigma^+$), and the exothermicity can also simply be transferred to kinetic energy of the reaction products. A measurement of the population of each of the different excited states was not available. The order of magnitude of the excitation probability of A$^2\Delta$ is estimated in \sref{sec:excitation} by comparing modeling results to experimental values.

\begin{table}
\caption{\label{DR_reactions_list}Dissociative recombination reactions for the methane family, with the labels used throughout this paper to refer to them. Also given are their exothermicities, taken from \cite{JR2002Methane}}.
\begin{indented}
\item[]\begin{tabular}{@{}llll}
\br
Label &Reaction&Exothermicity (eV)\\
\mr
CH$_4^+$ D.R. & $e + \mathrm{CH}_4^+ \rightarrow \mathrm{CH(*)} + \mathrm{H} + \mathrm{H}_2$ & 3.42 \\
CH$_3^+$ D.R. (a) & $e + \mathrm{CH}_3^+ \rightarrow \mathrm{CH(*)} + \mathrm{H}_2$ & 5.10 \\
CH$_3^+$ D.R. (b) & $e + \mathrm{CH}_3^+ \rightarrow \mathrm{CH} + \mathrm{H} + \mathrm{H}$ & 0.64 \\
CH$_2^+$ D.R. & $e + \mathrm{CH}_2^+ \rightarrow \mathrm{CH(*)} + \mathrm{H}$ & 6.00\\
\br
\end{tabular}
\end{indented}
\end{table}

%% file: puffing.tex
In various experiments, methane (CH$_4$) was injected into the magnetised hydrogen plasma beam of Pilot-PSI. The plasma beams used in this study typically had $T_e \sim 1$ eV, $n_e \sim 10^{20}$ m$^{-3}$ at the plasma axis with approximately 10 mm full width half maximum (FWHM) for both parameters. The magnetic field was $0.4$ T. CH emission was measured using an absolutely calibrated CCD camera equipped with a bandpass filter (peak transmission at 432.2 nm and a 2.0 nm FWHM Gaussian transmission function). Reference light intensities observed without CH puffing were subtracted from the observed light. The measured intensity was then multiplied by a factor 2.8 to correct for the fact that only part of the CH band passes the filter \cite{brezinsek2008CHpuff}. This way, line-integrated 2D profiles of CH emission can be conveniently measured. Deviations from these settings are noted with the individual experiments.

The total emission efficiency is characterised by the photon efficiency
\begin{equation}\label{eq:PIphot}
\Pi_{phot}=\frac{\phi^{CH}_{A \rightarrow X}\ [\mathrm{photons} \cdot \mathrm{s}^{-1}]}{\Gamma_{CH_4}\ [\mathrm{molecules} \cdot \mathrm{s}^{-1}]},
\end{equation}
where $\phi^{CH}_{A \rightarrow X}$ is the total photon flux per second and $\Gamma_{CH_4}$ is the number of injected methane molecules per second. By assuming that the dominant erosion product is CH$_4$, one can calculate the gross erosion flux in an erosion experiment by multiplying the measured CH A $\rightarrow$ X photon flux by $\Pi_{phot}$. For this reason, the photon efficiency is commonly used for calibrating erosion measurements \cite{westerhout2009erosion, whyte2001PISCES, nakano2002JT60U, brezinsek2004Review}. Under the ``corona assumption'' (a homogeneous low-density plasma), $\Pi_{phot}$ is determined only by the rate coefficients of hydrocarbon dissociation and of CH excitation. In such cases it can be obtained by 0-dimensional calculation, such as the HYDKIN toolbox \cite{HYDKIN}. But in reality, carbon redeposition and reflection as well as temperature gradients and transport losses in the plasma do influence the emission significantly. This makes detailed transport modeling necessary for CH emission and erosion quantification in tokamaks.

%% file: simulation.tex
The present study is a continuation of previous modeling efforts in Pilot-PSI \cite{borodin2010}. Methane test particles leave a puffing hole (\diameter ~0.6 mm) in the target at a thermal energy distribution corresponding to a source temperature of 700 $^\circ$C (approximately the target temperature), with a cosine angular distribution. Quickly after injection, they get ionised and dissociated predominantly through charge exchange and dissociative recombination. Reaction rates for methane breakdown come from {\cite{JR2002Methane}}. The modeling used a static background plasma, with plasma parameters from experimental Thomson scattering measurements.

Two photon emission channels are taken into account. Cross sections for CH excitation by electron impact were taken from \cite{celiberto2009}. Furthermore, every dissociative recombination event capable of producing excited CH was counted (keeping in mind that only a fraction of these will actually produce excited CH). In both cases, light emission is assumed to occur at the same point in space as the excitation, due to the short lifetime of the CH A-level. Due to the high electron densities, it appeared necessary to also take \textit{de-excitation} by inelastic electron impact collisions and by dissociation into account.

Though this study focuses on the quantitative description of photons from dissociative recombination, several improvements have been made in comparison with previous modeling. They will be outlined below. The effects of these modeling assumptions on simulation results are discussed in \sref{sec:discussion} along with an error analysis.

\subsection{De-excitation of CH (A $^2\Delta$)}\label{sec:quenching}
The radiative lifetime of the CH A $^2\Delta$ level is relatively short (0.53 $\mu$s \cite{becker1980CH}). Therefore, in low-density plasmas it is often assumed that excitation of a CH molecule is always followed by radiative decay. However, in high-density plasmas, one should not neglect the importance of de-excitation of CH by electron impact and by dissociation. The rate of electron de-excitation can be obtained from the excitation rate by using the principle of detailed balance \cite{FantzWunderlich2010}:
\begin{equation}
k_{A\rightarrow X} (\hat{T_e}) = k_{X\rightarrow A}(\hat{T_e}) \frac{g_X}{g_A} \exp \left(-\frac{\Delta E}{\hat{T_e}} \right)
\end{equation}
with $k_{X\rightarrow A}$ and $k_{A\rightarrow X}$ the excitation and de-excitation rates, $g$ the statistical weight of the given levels (where for CH, $g_A=g_X$), and $\Delta E$ the excitation energy. At higher electron densities, this de-excitation causes quenching of the photon emission.

There is more than one mechanism that can quench the CH A$\rightarrow$ X photon emission. Quenching also occurs if the rate of ionisation or dissociation is large enough to be comparable with the inverse radiative lifetime of CH. This possibility was taken into account, but found to be of relatively minor importance, as the rate of dissociation is much lower than that of electron impact de-excitation. The rate coefficient of the primary dissociation channel of CH (charge exchange with H$^+$) is $1.29 \cdot 10^{-15}$~m$^{-3}$ at T$_i = 1$~eV, resulting in an average lifetime of 8 $\mu$s, which is one order of magnitude above the radiative lifetime. It is also possible that electron impact causes transfer from the A$^2\Delta$ level to other excited states, such as the B$^2\Sigma^-$ or the C$^2\Sigma^+$ levels. In particular, it is known that the A$^2\Delta$, $\nu=0$ and the B$^2\Sigma^-$, $\nu=1$ are near-degenerate and collisional interconversion between these states can proceed efficiently \cite{Randall2000}. Since the population of the B$^2\Sigma^-$, $\nu=1$ level is typically much lower than that of the A$^2\Delta$, $\nu=0$ level, such collisional coupling should cause net reduction of the A$^2\Delta$, $\nu=0$ population. A complete collisional-radiative model for the CH radical would take this into account, but unfortunately no reaction rates are available for the electron-impact collisional interconversion between these levels. Therefore the quenching rate used herein should be seen as a lower limit of the actual quenching rate.

\begin{figure}[h]
\includegraphics[width=7cm]{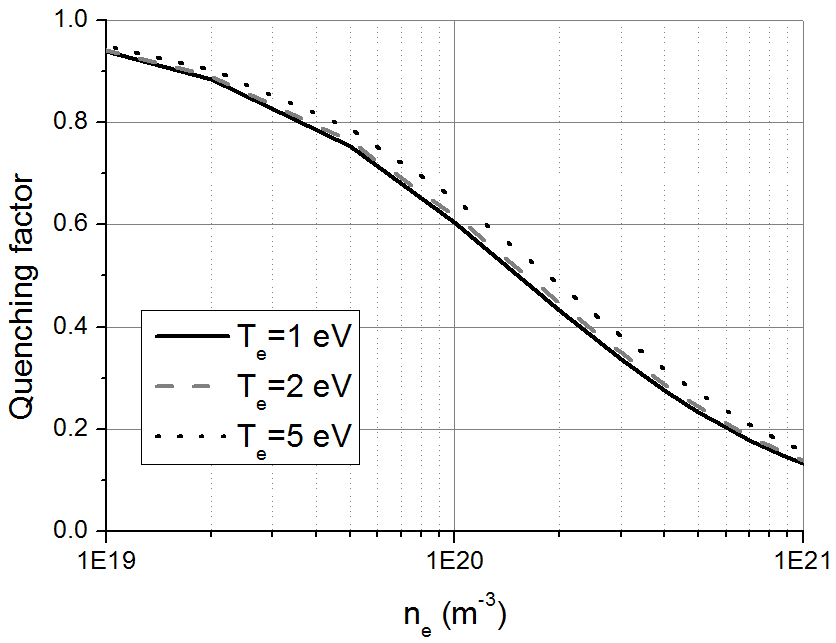}
\caption{\label{fig:quenchRate}Quenching factors for different electron temperatures.}
\end{figure}

The A-level is depleted by both spontaneous emission (occurring at a rate of the Einstein $A$-coefficient $A_{A\rightarrow X}$) and by de-excitation (occuring at a rate $n_e \cdot k_{A \rightarrow X}$). Therefore the effective photon emission rate is reduced by the same factor:
\begin{equation}
\mathrm{f}_{\mathrm{quench}}(T_e, n_e) = \frac{A_{A\rightarrow X}}{n_e \cdot k_{A\rightarrow X}+A_{A\rightarrow X}},
\end{equation}
such that f$_\mathrm{quench}=1$ indicates no quenching, and f$_\mathrm{quench}<1$ indicates that quenching reduces the effective photon emission. The quenching factor is plotted in \fref{fig:quenchRate}. Electron densities at the strike point of the ITER divertor are expected to be above 10$^{20}$ m$^{-3}$. In such conditions f$_\mathrm{quench}$ is significantly below 1, such that CH quenching by de-excitation should not be neglected.

\subsection{Pre-sheath treatment}\label{sec:presh}
Plasma ions are accelerated towards the sheath edge in the so-called pre-sheath region. The sheath itself is very thin ($\sim$ 10 $\mu$m), but the pre-sheath length is on the order of a cm. Typically, the decay length of the CH emission intensity is in the same order of magnitude. Therefore it is necessary to consider the acceleration and associated density drop in the pre-sheath. Due to the low electron temperatures in Pilot-PSI, ionisation inside the pre-sheath is negligible. Pilot-PSI has a flowing plasma, and collisions with neutral hydrogen are significant; this strongly influences the pre-sheath. In Pilot-PSI, the pre-sheath is formed by momentum loss due to collisions between plasma ions and the neutral hydrogen before the target. Such a ``collisional pre-sheath'' has different properties than the ionizing pre-sheath which is found in plasmas with higher electron temperatures {\cite{Riemann1991Presheath}}.

The thickness of the pre-sheath is determined by the mean free path $\lambda_{ni}$ with which incoming plasma ions lose their momentum to the neutrals before the target. Typically, the main momentum loss mechanism is charge exchange with H atoms reflected from the surface. Charge exchange with molecular hydrogen and elastic collisions are only of secondary importance. A typical value for $\lambda_{ni}$ of 7 mm has been found by fitting model and experiments {\cite{shumack2011Thesis}}. The corresponding H density $n_H$ can be found using the cross section $\sigma_{CX}$ for the dominating charge exchange reaction (H + H$^+$ $\rightarrow$ H$^+$ + H), which is 6 $\cdot$ 10$^{-19}$ m$^2$ \cite{HYDHEL}. From that, one finds $n_H = 1/(\lambda_{ni}*\sigma_{CX}) \approx 2 \cdot 10^{20}$ m$^{-3}$. The ion flow velocity far away from the target was taken from measurements \cite{vijvers2008PhysPlasmas} to be $4000$ m s$^{-1}$. The density drop and velocity profile corresponding to those parameters were implemented in these simulations. The resulting density drop between the Thomson scattering laser and the target is a factor of 3.4. This number is somewhat greater than the typically assumed pre-sheath density drop of a factor 2, because the friction between ions and neutrals is important in Pilot-PSI.

\subsection{Other modeling improvements}\label{sec:ModelImprove}
The Pilot-PSI plasma flow normally rotates around the plasma axis. Though maximum rotation velocities of over 10 km s$^{-1}$ were measured close to the plasma source {\cite{shumack2008Rotation}}, rotation velocities measured close to the target were significantly lower {\cite{shumack2011Thesis}}. In this work, the maximum rotation velocity was set at 3000 m s$^{-1}$ at $r=7.0$ mm away from the plasma axis; this value was taken from the experiment best matching conditions in this paper. The electric field that causes this plasma rotation was also taken into account. No big qualitative changes were found due to this rotation.

In Pilot-PSI the ion density is of the same order as the neutral density \cite{shumack2011diagnosing}. Therefore Coulomb friction is the dominant force affecting the test particles. For that reason collisions between the test particles and neutral hydrogen were neglected, speeding up the simulations. Because of the strong re-erosion caused by the high hydrogen flux to the target, carbon redeposition was neglected. All carbon molecules arriving at the target are assumed to reflect as CH$_4$.

%% file: results.tex
\subsection{Puffing into the side of the plasma beam}\label{sec:sidepuff}
Methane was injected perpendicularly through a nozzle into the side of the plasma beam at a rate of 3 sccm ($1.34 \cdot 10^{18}$ molecules s$^{-1}$). The injection nozzle (0.6 mm diameter) was located 25 mm away from the target and 17 mm away from the center of the beam. There, the axial gradients of the plasma parameters are small. This setup has two advantages. First, carbon recycling at the target is not so important here, due to the distance between the nozzle and the target. Second, the Thomson scattering laser that measures electron temperature and density is located at the same distance from the target as the CH$_4$ injection nozzle, thus enabling measurements of the plasma parameters at the location of injection.

\begin{figure}[h]
\includegraphics[width=8cm]{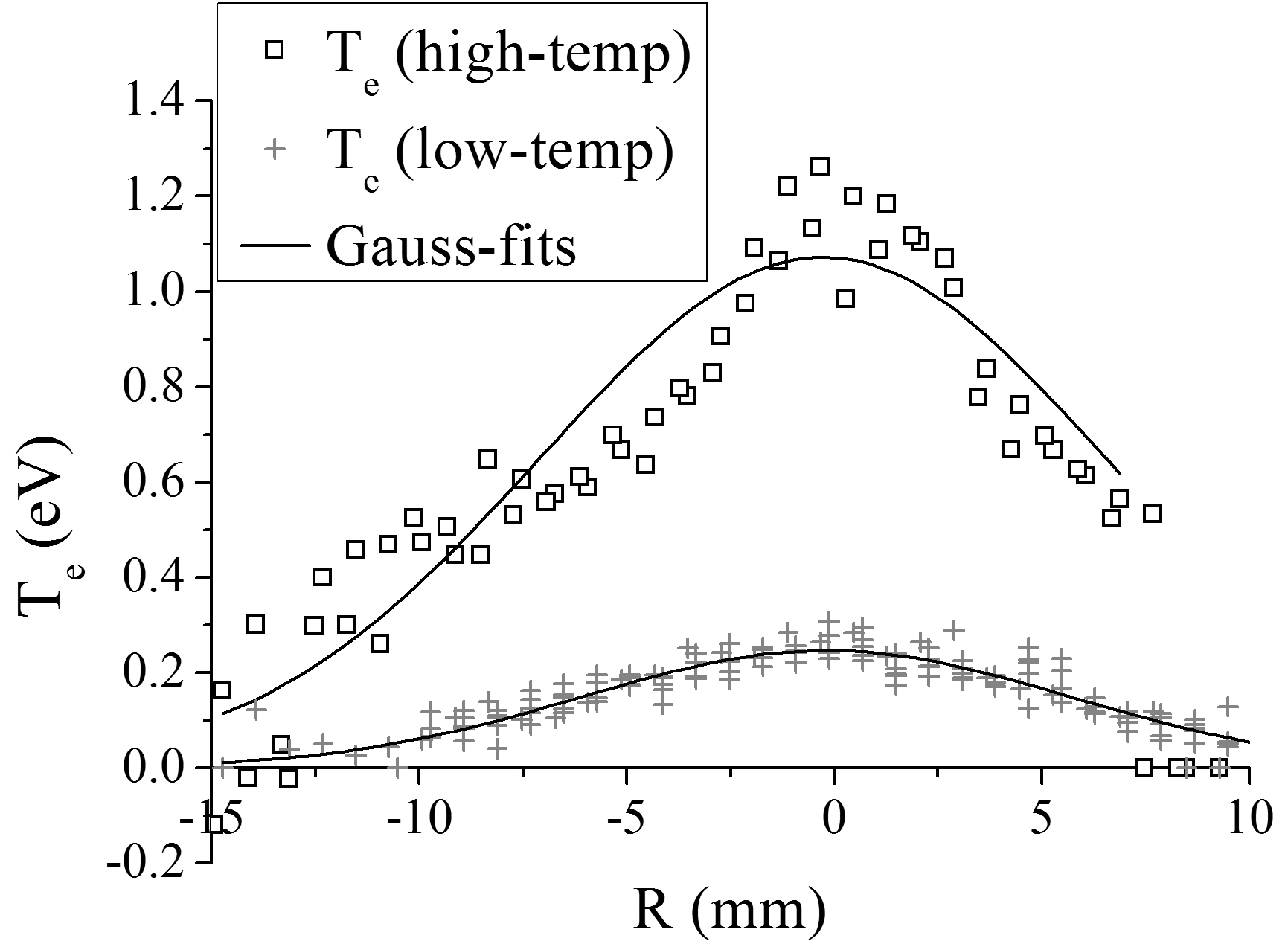}
\includegraphics[width=8cm]{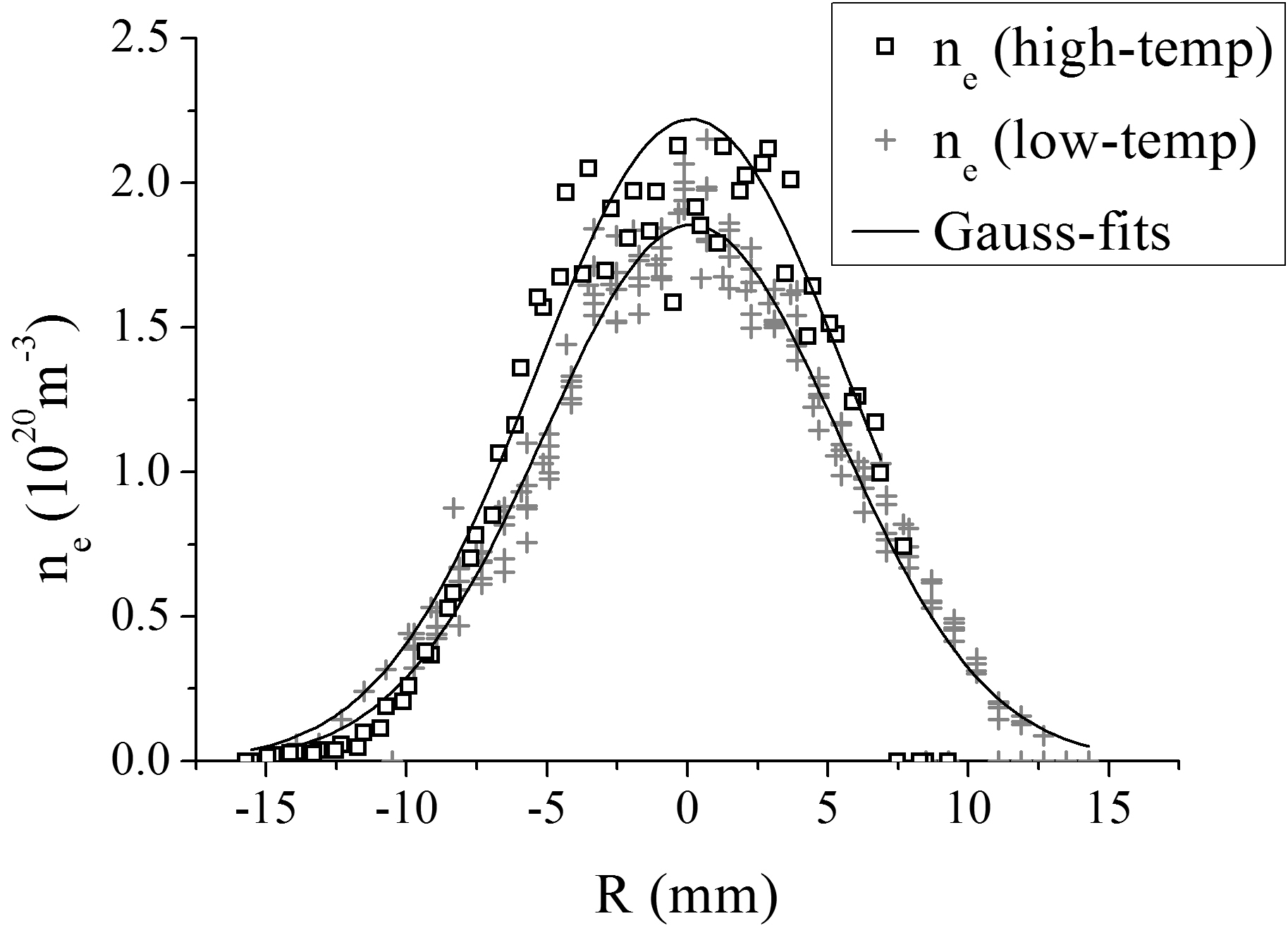}
\caption{\label{TS_sidepuff}Thomson scattering measurements for the side puffing experiment, and the Gaussian fits that were used for simulation input.}
\end{figure}

Both the measurements and the simulations were performed with two different electron temperatures, which we will call the ``high $T_e$'' and ``low $T_e$'' cases. Electron temperature and density, as measured by Thomson scattering, are shown in \fref{TS_sidepuff}.

\begin{figure}[h]
\caption{\label{fig:sidepuff_plumes}Experimental (left) and simulated (right) emission profiles during the side-puffing experiment. In the high-$T_e$ case, the modeled number of photons from both electron-impact (E.I.) excitation and dissociative recombination (D.R.) are shown. E.I. excitation is negligible in the low-$T_e$ case. The plots of D.R. events show the sum of all relevant D.R. channels. The small light plume close to the injection nozzle in the low T$_e$ experiment is due to reflection on the injection nozzle.}
\begin{tabular}{@{}lll}
\br
\includegraphics[width=8cm]{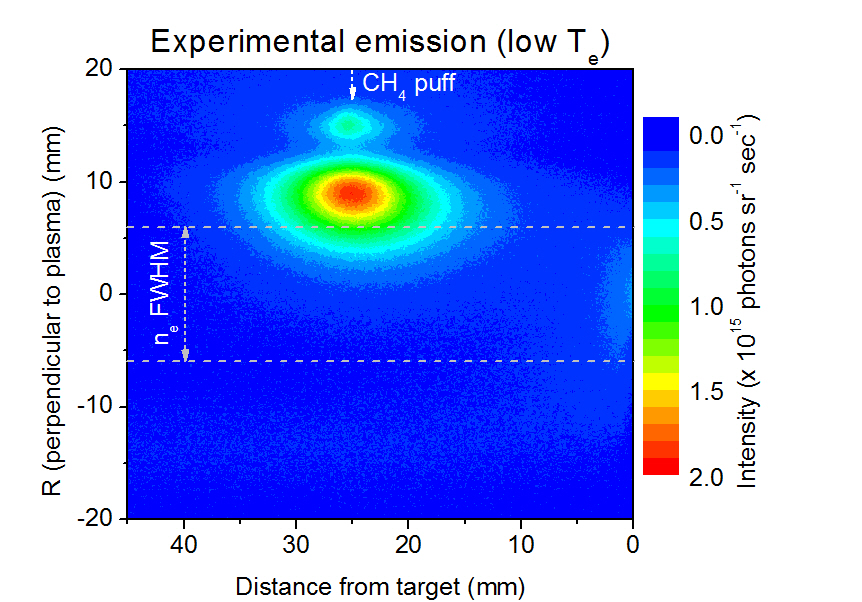} & \includegraphics[width=8cm]{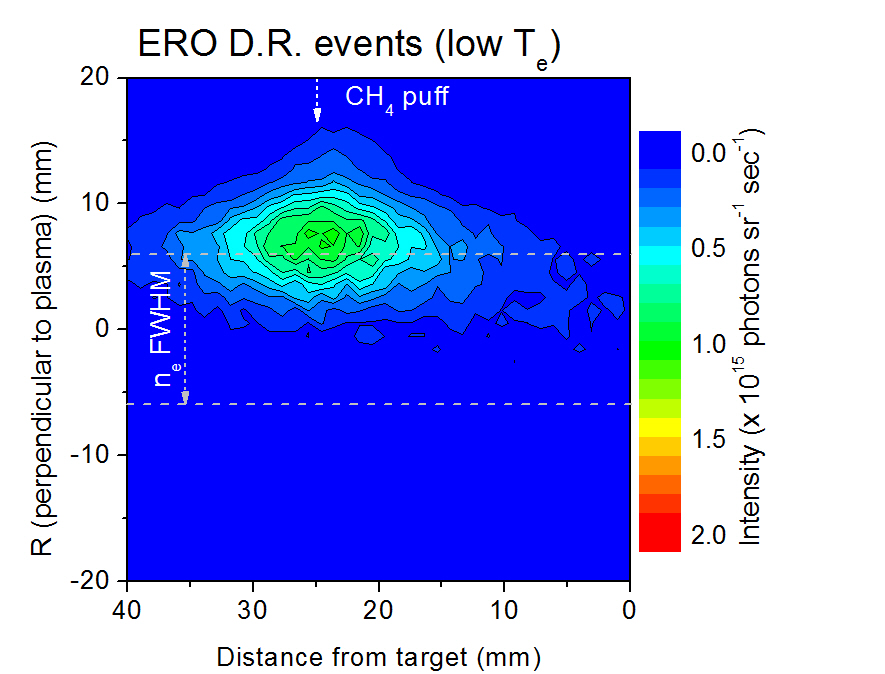} \\
\br
\multirow{2}{*}{\includegraphics[width=8cm]{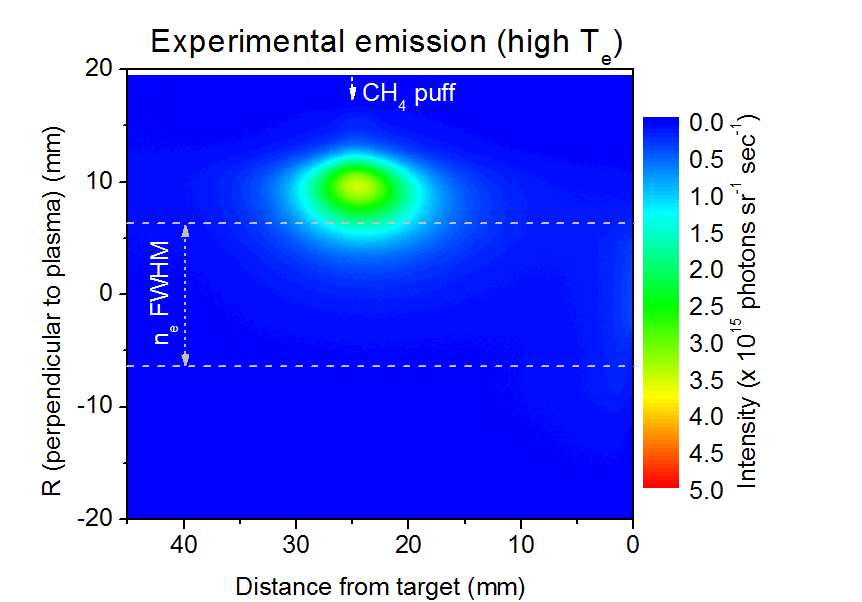}} &
\includegraphics[width=8cm]{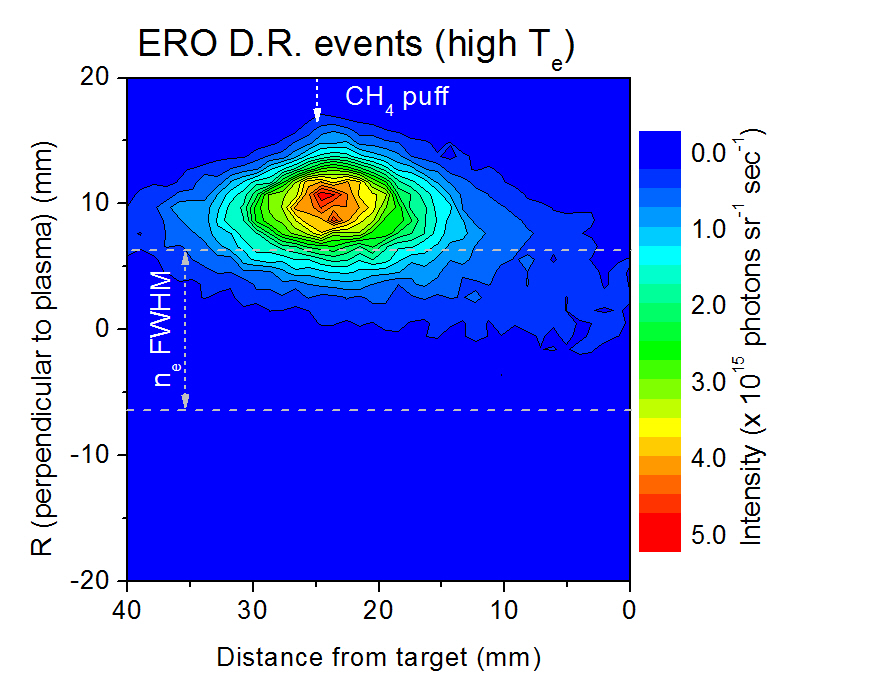} \\ & \includegraphics[width=8cm]{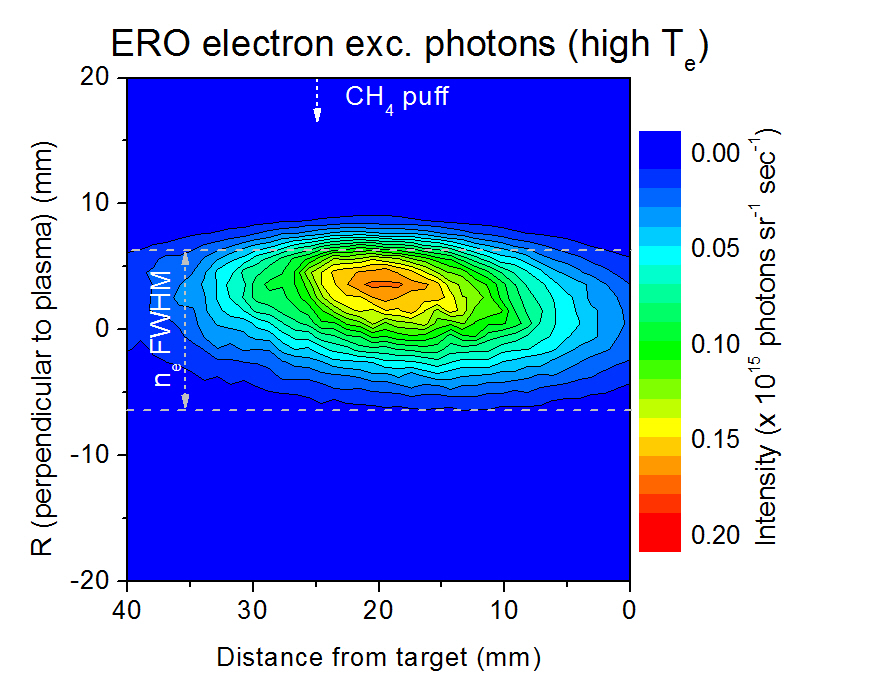} \\
\br
\end{tabular}
\end{figure}

Measured and simulated plumes are shown in \fref{fig:sidepuff_plumes}. We do not yet know which fraction of the D.R. events will produce a CH radical in the A$^2\Delta$ level (see \sref{sec:excitation}). Therefore, the D.R. ``photon count'' is an upper limit for the actual number of photons. It is clear from these results that the dissociative recombination produces a qualitatively different emission plume than the electron-impact (E.I.) excitation. The total photon yield from E.I. excitation is also far less than that from D.R., even in the experiment with higher T$_e$. The location of the simulated D.R. plume is in good agreement with the experimentally observed plume size, indicating that in both conditions electron-impact excitation is \textit{not} the dominant process. Note that this does not imply that the predicted electron-impact excitation is in mismatch with the experiment; it is simply not the dominant process under these conditions.

Another interesting observation is that, similar to earlier methane injection experiments from a different geometry \cite{westerhout2010}, the observed photon emission coefficients were almost independent of electron temperature.

\begin{table}[h]
\caption{\label{PIphot_sidepuff}Effective photon emission coefficients $\Pi_{phot}$ as measured in the side-puffing experiment and modeled values from different processes.}
\begin{indented}
\item[]\begin{tabular}{@{}lll}
\br
 &$\Pi_{phot}$ (low-$T_e$ case)&$\Pi_{phot}$ (high-$T_e$ case)\\
\mr
Experiment & 0.038 & 0.033 \\
E.I. excitation & $<$0.001 & 0.009 \\
CH$_2^+$ D.R. & 0.003 & 0.019 \\
CH$_3^+$ D.R. & 0.012 & 0.032\\
CH$_4^+$ D.R. & 0.025 & 0.101 \\
\br
\end{tabular}
\end{indented}
\end{table}

The total effective photon emission coefficients $\Pi_{phot}$ (\eref{eq:PIphot}) are given in \tref{PIphot_sidepuff}. Of the D.R. channels, the CH$_4^+$ D.R. ($e + \mathrm{CH}_4^+ \rightarrow \mathrm{CH(*)} + \mathrm{H} + \mathrm{H}_2$) occurs by far most frequently. This does not necessarily imply that the reaction will be dominant in actually exciting the A-level, since the probability of yielding an excited radical is not given, as will be discussed in \sref{sec:excitation}. However, the sum of all other emission channels is not enough to explain the observed photon emission yields. That means that the fraction of CH$_4^+$ D.R. that yields an excited radical should be significant in order to explain the experimental $\Pi_{phot}$ at low T$_e$.

\subsection{Puffing into the center of the plasma beam}\label{puffCenter}
\begin{figure}
\includegraphics[width=6cm]{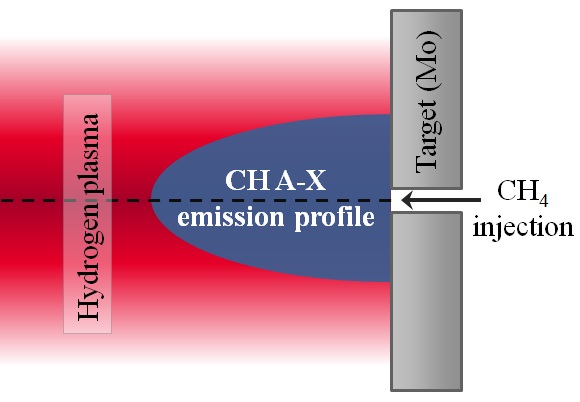}
\caption{\label{fig:ill_puff}}Illustration of the experiments of \sref{puffCenter} and \sref{sec:neScan}.
\end{figure}

In the following two series of experiments, methane was injected directly into the center of the plasma beam, through a hole in the target. This geometry is illustrated in \fref{fig:ill_puff}. 

\begin{figure}[h]
\includegraphics[width=8cm]{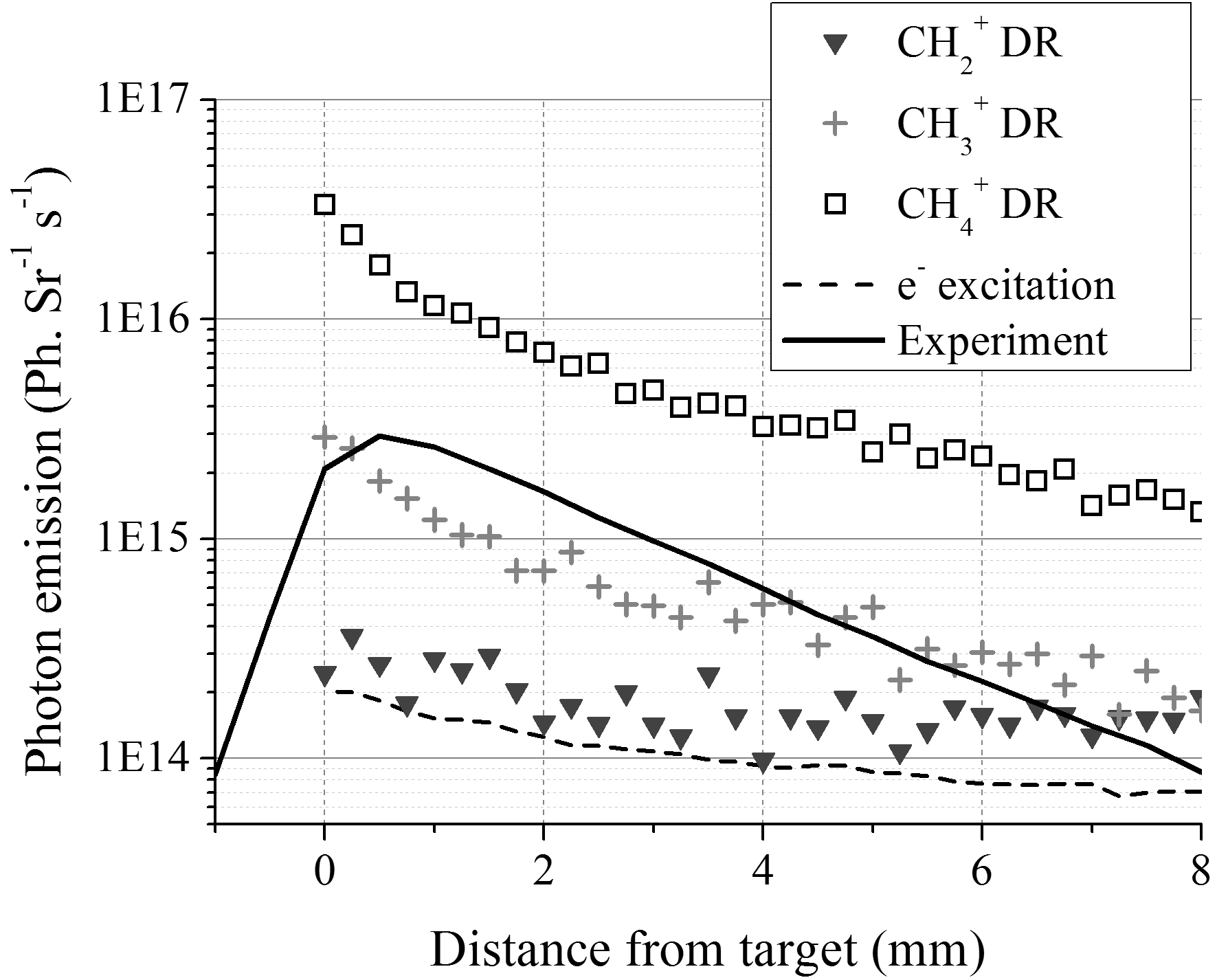}
\caption{\label{fig:targetpuff_profile}Photon emission profiles in the center of the plasma at $T_e$=1.25 eV. Experimental values are from \cite{westerhout2010}.}
\end{figure}

Profiles of the (line-integrated) CH emission in the center of the plasma beam are shown in \fref{fig:targetpuff_profile} together with simulation results. The profile is taken along the dashed line in \fref{fig:ill_puff}. On a logarithmic scale, the light emission quickly drops after several millimeters. From the decay length of the various excitation channels, one can see that the dissociative recombination of CH$_3^+$ gives a good match with experiment. Also the DR of CH$_4^+$ will match experiment, if weighted by an appropriate excitation probability. CH$_2^+$ DR, by itself, is not enough to explain experimental photon emission.

\begin{figure}[h]
\includegraphics[width=12cm]{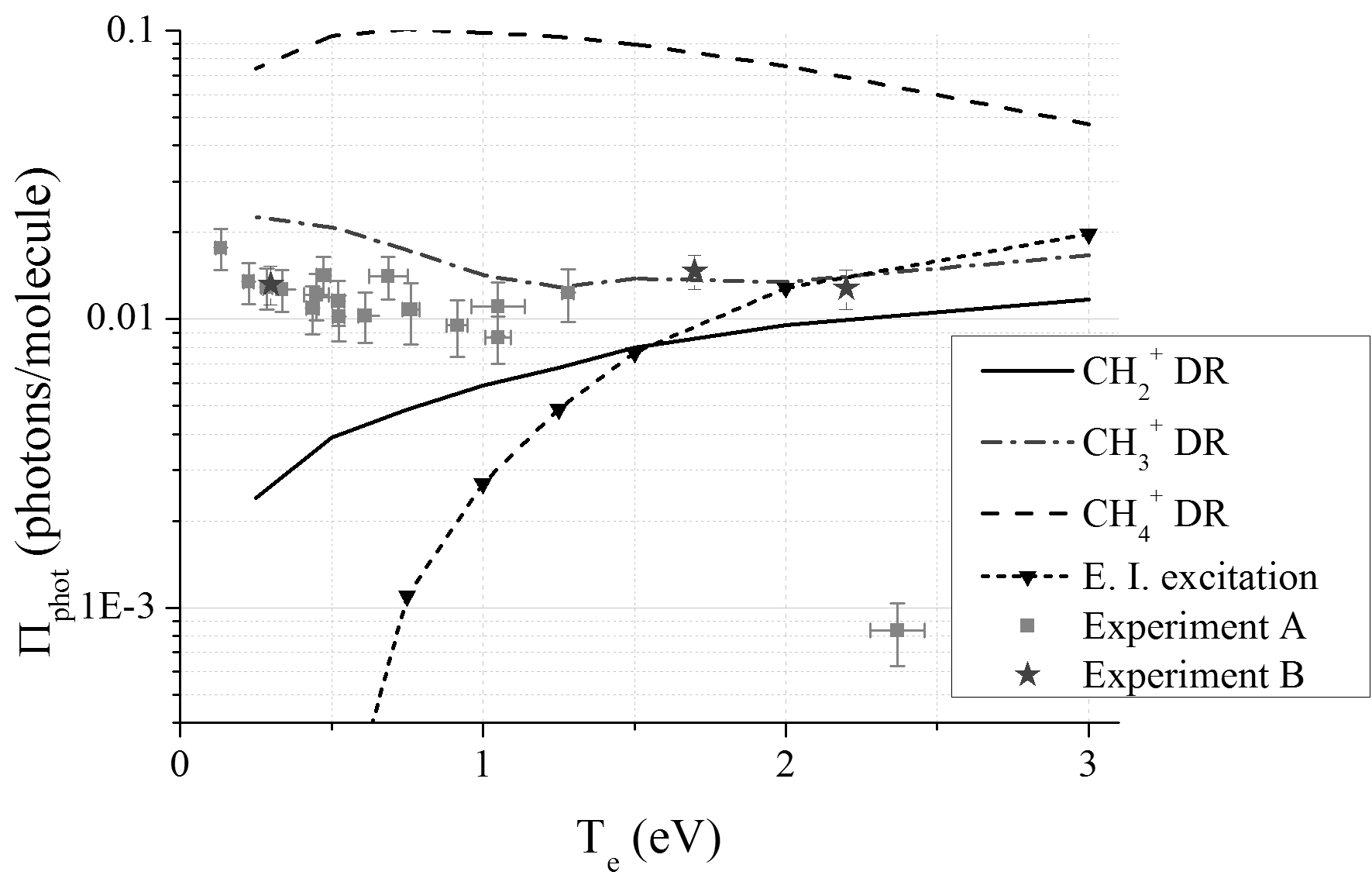}
\caption{\label{targetpuff_PIphot}$\Pi_{phot}$ from experiment and from simulation of the various CH excitation channels, as a function of $T_e$. Details can be found in the main text.}
\end{figure}

We would now like to establish the T$_e$-dependence of the various photon emission processes. The absolute CH A-X photon emission was measured during scans of T$_e$. These experiments are described in detail elsewhere: ``experiment A'' was performed using a spectrometer \cite{westerhout2010}, and ``experiment B'' was performed using a CCD camera \cite{vegt2011Puffing}. Both the spectrometer and the CCD camera were absolutely calibrated. Therefore, both experimental results can be compared with each other and with simulation results. Results are shown in \fref{targetpuff_PIphot}. The photon yield of each of the D.R. processes (table \ref{DR_reactions_list}), as well as from electron impact excitation is shown separately. It should be noted that the experimental point at 2.4 eV had a higher magnetic field (0.8 T, rather than 0.4 T in the other points); this led to a relatively high electron density. That might explain the low photon yield in that experiment. Nevertheless, the magnitude of this discrepancy is unexpected, and, for the time being, unexplained. 

Below 1.5 eV, E.I. excitation drops steeply due to the threshold character of that process. In that regime, the dissociative recombination channels are clearly necessary to explain the photon yield. Like in section \ref{sec:sidepuff}, CH$_4^+$ dissociative recombination is the most frequently occurring D.R. channel. By just counting CH$_4^+$ D.R. events, we overestimate the experimental $\Pi_{phot}$. This suggests that only a fraction of these D.R. events actually produces a photon. Due to the discrepancy between experimental results at $T_e > 2$ eV, it is not yet possible to give a precise temperature threshold at which D.R. becomes dominant. We can, however, conclude that certainly below T$_e = 1.5$ eV, D.R. is necessary (and sufficient) to explain the intensity of CH emission in these experiments.

\subsection{Electron density scan, importance of CH de-excitation}\label{sec:neScan}
A scan of the electron density was performed both experimentally and in modeling. Such a scan can shed light on the role of de-excitation of the CH emission (see section \ref{sec:quenching}). This quenching increases as the electron density goes up, as shown in figure \ref{fig:quenchRate}. In these experiments, methane was puffed through the target into the center of the plasma. As in \cite{westerhout2010}, the photon yield was measured with a spectrometer; however, it has not been absolutely calibrated. The magnetic field was set to 0.8 T to reach high electron densities. T$_e$ varied between 0.5 and 0.9 eV; each individual simulation used the T$_e$ and n$_e$ that were measured in the corresponding experimental point.

\begin{figure}[h]
\includegraphics[width=10cm]{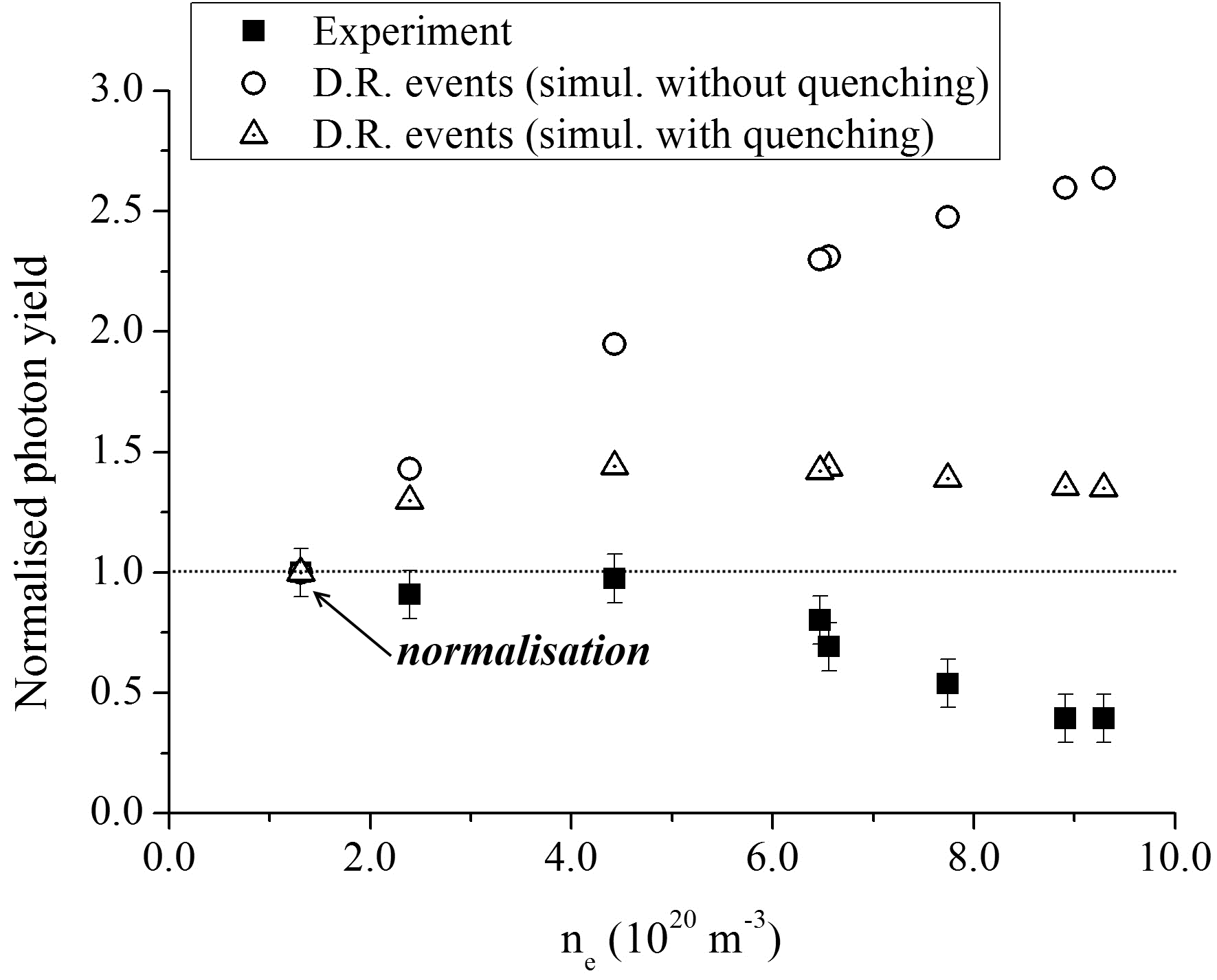}
\caption{\label{fig:neScan}Normalised photon yield from experiment, and D.R. photons from ERO simulation at various densities. D.R. photons are summed from all relevant channels in table \ref{DR_reactions_list}. Densities on the x-axis are those measured at the location of the Thomson scattering diagnostic.}
\end{figure}

In one simulation series the effect of de-excitation was taken into account, in another series it was neglected. Both were then compared with experiment. Because T$_e < 1$ eV, the contribution from E.I. excitation is relatively small. Hence, only the sum of all dissociative recombination events was compared with experimental values. Results are shown in fig. \ref{fig:neScan}. Because the spectrometer was not absolutely calibrated, all results are normalised to 1 at the lowest density point.

It should be noted that the density was measured by a Thomson scattering diagnostic at 25 mm before the target. As described in \sref{sec:presh} the density drops in the final cm before the target, due to friction with neutrals in the pre-sheath. 

At the lowest electron density (n$_e=1.3*10^{20}$ m$^{-3}$), the calculated emission rate is already reduced by a factor 1.4 due to the quenching (though this is hidden by the normalisation). At the highest electron density (n$_e=9.3*10^{20}$ m$^{-3}$), the reduction by quenching is a factor 2.8. Experimentally, it was observed that the photon yield goes down as the electron density increases, and this was attributed to de-excitation. However, the simulations without the inclusion of de-excitation give exactly the opposite result: the photon yield goes up with increasing density. The explanation for this modeling result is that with a higher electron density, hydrocarbon molecules are dissociated more rapidly, significantly reducing the loss fraction of hydrocarbon molecules that leave the plasma before they are dissociated to CH. The simulation series that includes CH de-excitation is considerably closer to experiment than the series without de-excitation.

Even though the trend of a decreasing photon yield which increasing density is not yet fully reproduced, these results do underline the significance of de-excitation of CH in high-density plasmas. Since collisional interconversion between the A$^2\Delta$ level and other excited levels could not be taken into account, the quenching rate used in these calculations is a lower limit of the actual quenching rate (see \sref{sec:quenching}). That is a possible explanation for the discrepancy between the experiment and simulation.

%% file: exc_prob.tex
In all of the simulations, CH$_4^+$ D.R. is the most frequently occurring D.R. channel. This can be explained by examining the reaction database that is used by ERO \cite{JR2002Methane}. From these reaction rates it follows that many of the methane molecules entering the plasma are quickly ionised to CH$_4^+$ by charge exchange with H$^+$. CH$_4^+$ then quickly recombines to a variety of products, including (25 \%) CH. The other D.R. channels that can excite the CH radical are CH$_2^+$ D.R. and CH$_3^+$ D.R.

From these results, it is not trivial to estimate the fraction of each D.R. channel that excites the CH level. The electron temperature scan in \fref{targetpuff_PIphot} suggests that in the case of CH$_4^+$ D.R. this fraction should be on the order of 10\% to match the experimentally observed photon flux. However, looking at \tref{PIphot_sidepuff}, one estimates a somewhat larger value for this fraction. Either way, it does seem that excitation of the CH A$^2 \Delta$ level is relatively common during the dissociative recombination of CH$_x^+$ ions. Both experiments suggest that these probabilities are at least of the order of 10\%. Judging from \fref{targetpuff_PIphot}, it seems that the effect of E.I. excitation does become significant starting from electron temperatures around 1.5-2.5 eV. Thus, even though D.R. excitation of CH probably does generate \textit{some} photons in the CH$_4$ puffs in detached plasma at JET (T$_e \sim 2$ eV) \cite{brezinsek2009JET} and DIII-D (T$_e = 2.0$ eV) \cite{McLean2007DIID}, it might be a marginal process. Further research is needed to conclusively answer this question.

Unfortunately, the branching ratio of the dissociative recombination of CH$_4^+$ into various hydrocarbons CH$_x$ has never been directly measured. Instead, these branching ratios are extrapolated from measurements on D.R. of CH$_3^+$ \cite{vejby-christensen1997CH3plus} and CH$_5^+$ \cite{semaniak1998CH5plus}. Because of the apparently important contribution of this channel at low T$_e$, experimental confirmation of this branching ratio would be beneficial to modeling efforts.

The fact that our simulations consistently give the D.R. of CH$_4^+$ to be the most frequently occuring process that can excite CH complicates the benchmarking of spectroscopy during erosion experiments with methane injection experiments at low T$_e$. In models of chemical erosion, CH$_3$ is often assumed to be an important eroded hydrocarbon species \cite{mech1997JAP}. That species does not readily form CH$_4^+$ in the plasma. This means that the dominant excitation mechanism may be different in erosion experiments and in injection experiments, complicating analysis of erosion experiments.

%% file: discussion.tex
Aside from the general measurement error in the underlying reaction rate databases, there are a number of uncertainties specific to these simulations; the major ones will be summarised briefly here.

\begin{itemize}
\item The properties of the pre-sheath (discussed briefly in section \ref{sec:presh}) have a significant influence on the results of ERO calculation, which is greater than the effect of other assumptions on plasma parameters (section \ref{sec:ModelImprove}). The drop in electron density close to the target causes the impurity penetration depth to increase to a greater value than the experimental observation. However, varying the pre-sheath parameters within a realistic range does not have such a great influence on the integrated photon emission yield; this remains within a factor two throughout a parameter scan. For that reason, only integrated photon yields are compared here. The only exception is the puff into the side of the beam, which is not so strongly affected by the pre-sheath. The exact pre-sheath density profile and plasma acceleration remains an interesting subject for future study.

\item With puff-rates of up to 3 sccm, the injection flux density right behind the target can be comparable to the hydrogen flux. For instance, when puffing into the center of the plasma through a 0.6 mm nozzle, the hydrocarbon flux density right behind the nozzle is $1 \cdot 10^{24}$ m$^{-2}$ s$^{-1}$, which is about the same as the incoming hydrogen ion flux. Furthermore, cauliflower-shaped carbon microparticles have been found on the target both in carbon erosion and puffing experiments in Pilot-PSI \cite{bystrov2011Carbon, bystrov2011, wright2010}. If these are formed in the plasma, it means there has to be formation of higher hydrocarbons (C$_x$H$_y$, $x > 1$). Therefore one should investigate whether the test particle assumption (i.e. no collisions between impurity molecules, and no influence of the impurity on the plasma) is still valid. If not, the influence of the hydrocarbon injection on local plasma parameters would have to be taken into account, as was done earlier for injection experiments in TEXTOR \cite{ding2011}.

A scan of the injection rate from 0.6 to 3 sccm has been performed \cite{vegt2011Puffing}. There, no obvious disturbance of the plasma, change in the CH penetration depth, or non-linearity in the total photon emission yield was observed. Thus, it seems that the central assumption of ERO is reasonable. The explanation for that may be that hydrocarbon molecules quickly spread radially through the plasma beam, so that their density is only a significant fraction of the plasma density in a very small area behind the injection nozzle.

\item Unlike previous calculations \cite{borodin2010}, all hydrocarbon radicals and molecular ions arriving at the target were assumed to be reflected. The reason for this is the high hydrogen flux in Pilot-PSI, which causes strong re-erosion of the carbon deposition; normally, only a small fraction of the injected carbon is retained on the metallic target. Reference simulations were also performed with the opposite extreme assumption: all carbon molecules arriving at the target are permanently deposited. That extreme assumption typically reduced the calculated total photon yield by approximately a factor 2 compared to the simulations in this paper. The penetration depth of carbon in the plasma is not significantly affected. We conclude that the uncertainties about sticking do not affect qualitative results.
\end{itemize}

%% file: conclusions.tex
Many updates have been made in simulations of methane puffing experiments in Pilot-PSI. Most importantly, the contribution of dissociative recombination to the production of excited CH has been included in the modeling. By comparing simulations to a series of both new and existing CH$_4$ injection experiments, it was then confirmed that this process dominates at electron temperatures below 1.5 eV. These results suggest that the fraction of D.R. events that actually produce an excited CH radical is at least of the order of 10\%.

De-excitation of CH* by electrons was included in the modeling, and an experimental scan of electron density was performed. It was demonstrated that at higher electron densities, such as expected in the ITER divertor, de-excitation certainly cannot be neglected. Inclusion of photon emission quenching by electron-impact de-excitation in the modeling greatly improved the agreement between modeling and experiment in a scan of n$_e$.

In present tokamaks, electron temperatures near the plasma-facing surface are typically much greater than 3 eV. Under such conditions, electron-impact processes are very efficient in the dissociation of hydrocarbon ions. In simulations of such plasmas, good agreement is typically obtained between measured and predicted photon emission \cite{ding2009}, and the photon production by dissociative recombination can be neglected. However, this is clearly not the case for low-temperature, ITER divertor-relevant plasmas.

These observations shed new light on the interpretation of CH photon emission measurements. Finally, they could also be of relevance for low-temperature hydrocarbon laboratory plasmas experiments, which also often employ CH spectroscopy.

%% file: refs.tex
% OMITTED REFERENCES (not relevant enough):
% =========================================
%
%\bibitem{kirschner2004EnhErosion} Kirschner A, Wienhold P, Philipps V, Coad J P, Huber A, Samm U, JET EFDA Contributors 2004 {\it J. Nucl. Mater.} {\bf 328} 62-66
%\bibitem{kirschner2003JET} Kirschner A, Brooks J N, Philipps V, Coad J P and Contributors to the EFDA-JET Workprogramme 2003 {\it Plasma Phys. Control. Fusion} {\bf 45} 309-319
%\bibitem{AhoMantila2010JNM} Aho-Mantila L, Wischmeier M, Krieger K, Rohde V, M\"uller H W, Coster D P, Groth M, Kirschner A, Neu R, Potzel S, Sieglin B, Wolfrum E, The ASDEX Upgrade Team 2010 {\it J. Nucl. Mater.}, doi:10.1016/j.jnucmat.2010.10.080 (in press)
%\bibitem{JR2010Update} Reiter D and Janev R K 2010 {\it Contrib. Plasma Phys.} {\bf 50}:10 986-1013
%\bibitem{EL1987database} Ehrhardt A B and Langer W D 1987 {\it ``Collisional Processes of Hydrocarbons in Hydrogen Plasmas''}, Report PPPL-2477, Princeton Plasma Physics Laboratory, Princeton, N.J., USA

%% file: 2011_CH_emission_paper.bbl
\begin{thebibliography}{10}
\bibitem{roth2005} Roth J \etal 2005 {\it J. Nucl. Mater.} {\bf 337--339} 970
\bibitem{Federici2003JNM_B2EIRENE} Federici G \etal 2003 {\it J. Nucl. Mater.} {\bf 313--316} 11--22
\bibitem{Federici2001Review} Federici G \etal 2001 {\it Nucl. Fus.} {\bf 41} 1967
\bibitem{temmerman2009} Doerner R P, Baldwin M J, De Temmerman G, Hanna J, Nishijima J, Roth J, Schmid K, Tynan G R and Umstadter K 2009 {\it Nucl. Fusion} {\bf 49} 035002
\bibitem{roth2009PWIissues} Roth J \etal 2009 {\it J. Nucl. Mater.} {\bf 390--391} 1--9
\bibitem{westerhout2010} Westerhout J, Borodin D, Brezinsek S, Lopes Cardozo N J, Rapp, J, Schram D C and Van Rooij G J 2010 {\it Nucl. Fus.} {\bf 50} 095003
\bibitem{vanrooij2007_Pilot} Van Rooij G J \etal 2007 {\it Appl. Phys. Lett.} {\bf 90} 121501
\bibitem{beulens1994Emission} Beulens J J, Gastineau C, Guerrassimov N, Koulidiati J and Schram D C 1994 {\it Plasma Chemistry and Plasma Processing} {\bf 14}:1 15--42
\bibitem{kobayashi2003RotPop} Kobayashi H, Kado S, Xiao B and Tanaka S 2003 {\it Jpn. J. Appl. Phys} {\bf 42} 1776--1787
\bibitem{avtaeva2007} Avtaeva S V, Lapochkina T M 2007 {\it Plasma Phys. Rep.} {\bf 33}:9 774--785
\bibitem{dorier2009DRInSpraying} Dorier J L, Guittienne Ph, Hollenstein Ch, Gindrat M and Refke A 2009 {\it Surface and Coatings Technology} {\bf 203}:15 2125--2130
\bibitem{kirschner2007} Kirschner A, Borodin D, Droste S, Philipps V, Samm U, Federici G, Kukushkin A and Loarte A 2007 {\it J. Nucl. Mater.} {\bf 363--365} 91--95
\bibitem{kirschner2009JNM} Kirschner A, Borodin D, Philipps V, Samm U, Ding R, Schmid K, Roth J, Kukushkin A, Federici G and Loarte A 2009 {\it J. Nucl. Mater.} {\bf 390--391} 152--155
\bibitem{kirschner2009PhysScr} Kirschner A, Ohya K, Borodin D, Ding R, Matveev D, Philipps V and Samm U 2009 {\it Phys. Scr.} {\bf T138} 014011
\bibitem{kirschner2000} Kirschner A, Philipps V, Winter J and K\"ogler U 2000 {\it Nucl. Fus.} {\bf 40}:5 989
\bibitem{westerhout2007PhysScr} Westerhout J \etal 2007 {\it Phys. Scr.} {\bf T128} 18
\bibitem{borodin2010} Borodin D \etal 2010 {\it Contrib. Plasma Phys.} {\bf 50}:3--5 432--438
\bibitem{reiter2009} Reiter D, K{\"u}ppers B and Janev R K 2009 {\it Phys. Scr.} {\bf T138} 014014
\bibitem{JR2002Methane} Janev R K and Reiter D 2002 {\it Phys. Plasmas} {\bf 9} 4071
\bibitem{brezinsek2008CHpuff} Brezinsek S \etal 2007 {\it J. Nucl. Mater.} {\bf 363--365} 1119--1128
\bibitem{westerhout2009erosion} Westerhout J \etal 2009 {\it Phys. Scr.} {\bf T138} 014017
\bibitem{whyte2001PISCES} Whyte D G, Tynan G R, Doerner R P and Brooks J N 2001 {\it Nucl. Fusion} {\bf 41}:1 47--62
\bibitem{nakano2002JT60U} Nakano T, Kubo H, Higashijima S, Asakura N, Takenaga H, Sugie T and Itami K 2002 {\it Nucl. Fusion} {\bf 42} 686--696
\bibitem{brezinsek2004Review} Brezinsek S \etal 2004 {\it Phys. Scr.} {\bf 111} 42--48
\bibitem{HYDKIN} Reiter D, K\"uppers B 2008 http://www.hydkin.de
\bibitem{celiberto2009} Celiberto R, Janev R K and Reiter D 2009 {\it Plasma Phys. Control. Fusion} {\bf 51} 085012
\bibitem{becker1980CH} Becker K H, Brenig H H and Tatarczyk T 1980 {\it Chem. Phys. Lett.} {\bf 71}:2 242--245
\bibitem{FantzWunderlich2010} Fantz U and W\"underlich D 2010 {\it AIP Conf. Proc.} {\bf 1344} 204--216
\bibitem{Randall2000} Randall C J, Murray C and McKendrick K G 2000 {\it Phys. Chem. Chem. Phys.} {\bf 2} 461--471
\bibitem{Riemann1991Presheath} Riemann K U 1991 {\it J. Phys. D: Appl. Phys.} {\bf 24} 493--518
\bibitem{shumack2011Thesis} Shumack A E 2011 {\it The influence of electric fields and neutral particles on the plasma sheath at ITER divertor conditions}, PhD Thesis, pp. 117, Technical University Eindhoven
\bibitem{HYDHEL} Janev R K, Langer W D, Evans Jr. K, Post Jr. D E 1987 {\it Elementary Processes in Hydrogen-Helium Plasmas} ({\it Springer Series on Atoms and Plasmas} vol~4) (Berlin: Springer-Verlag)
\bibitem{vijvers2008PhysPlasmas} Vijvers W A J, Van Gils V A J, Goedheer W J, Van der Meiden H J, Schram D C, Veremiyenko V P, Westerhout J, Lopes Cardozo N J, Van Rooij G J 2008 {\it Phys. Plasmas} {\bf 15} 093507
\bibitem{shumack2008Rotation} Shumack A E, Veremiyenko V P, Schram D C, de Blank H J, Goedheer W J, van der Meijden H J, Vijvers W A J, Westerhout J, Lopes Cardozo N J and van Rooij G J 2008 {\it Phys. Rev. E} {\bf 78} 046405
\bibitem{shumack2011diagnosing} Shumack A E, Schram D C, Biesheuvel J, Goedheer W J and van Rooij G J 2011 {\it Phys. Rev. E} {\bf 83} 036402
\bibitem{vegt2011Puffing} Bystrov K, Van der Vegt L B, Van Swaaij G A, Zaharia T, Kuang Y, Goedheer W J and De Temmerman G 2012, submitted to {\it J. Nucl. Mater.}.
\bibitem{brezinsek2009JET} Brezinsek S \etal 2009 {\it J. Nucl. Mater.} {\bf 390--391} 267--273
\bibitem{McLean2007DIID} McLean A G \etal 2007 {\it J. Nucl. Mater.} {\bf 363--365} 86--90
\bibitem{vejby-christensen1997CH3plus} Vejby-Christensen L, Andersen L H, Heber O, Kella D, Pedersen H B, Schmidt H T, Zajfman D 1997 {\it Astrophys. J.} {\bf 483}:1 531--540
\bibitem{semaniak1998CH5plus} Semaniak J \etal 1998 {\it Astrophys. J.} {\bf 498}:2 886--895
\bibitem{mech1997JAP} Mech B V, Haasz A A and Davis J W 1998 {\it J. Appl. Phys.} {\bf 84}:3 1655--1669
\bibitem{bystrov2011Carbon} Bystrov K, Arnas C, Marot L, Van der Vegt L, Westerhout J, Van Rooij G J, De Temmerman G 2011. Submitted to {\it Carbon}
\bibitem{bystrov2011} Bystrov K, Westerhout J, Matveeva M, Litnovsky A, Marot L, Zoethout E, De Temmerman G 2011, {\it J. Nucl. Mater.} {\bf 415}:1 S149--S152
\bibitem{wright2010} Wright G M \etal 2010 {\it J. Nucl. Mater.} {\bf 396}:2--3 176
\bibitem{ding2011} Ding R \etal 2011 {\it J. Nucl. Mater.} {\bf 415}:1 S270--S273
\bibitem{ding2009} Ding R, Kirschner A, Borodin D, Brezinsek S, Pospieszczyk A, Schmitz O, Philipps V, Samm U, Chen J and Li J 2009 {\it Plasma Phys. Control. Fusion} {\bf 51} 055019
\end{thebibliography}
